\documentclass[10pt,twocolumn,brazil,english,amsmath,amssymb, aps]{revtex4-2}
\usepackage[T1]{fontenc}
\usepackage[utf8]{inputenc}
\usepackage{babel}
\usepackage{array}
\usepackage{multirow}
\usepackage{amsmath}
\usepackage{amssymb}
\usepackage{graphicx}
\PassOptionsToPackage{normalem}{ulem}
\usepackage{ulem}
\usepackage{color}
 \usepackage{hyperref}

\makeatletter

\providecommand{\tabularnewline}{\\}

\makeatother

\begin{document}
\title{Implications of a high growth index on the variation of $G$}
\author{Ícaro B. S. Cortês }
\email{icaro.cortes.710@ufrn.edu.br}

\affiliation{Departamento de Engenharia de Computação e Automação, Centro de Tecnologia,
Universidade Federal do Rio Grande do Norte, Natal, Rio Grande do
Norte, Brazil.}
\author{Léo G. Medeiros}
\affiliation{Escola de Ciências e Tecnologia, Universidade Federal do Rio Grande
do Norte, Natal, Rio Grande do Norte, Brazil.}
\author{Ronaldo C. Batista}
\affiliation{Escola de Ciências e Tecnologia, Universidade Federal do Rio Grande
do Norte, Natal, Rio Grande do Norte, Brazil.}
\begin{abstract}
A recent determination of the growth index indicates a value significantly higher than the $\Lambda$CDM prediction, suggesting that alternative scenarios to $\Lambda$CDM may be required. In this work, we investigate whether a time-varying Newton’s constant, $G_N$, can account for such a high growth index, $\gamma=0.633 ^{+0.024}_{-0.025}$. Adopting a phenomenological approach, we study two parametrizations of the effective gravitational coupling, $G_{\rm eff}$, one based on a Taylor expansion and another linked to the energy density parameter of Dark Energy. We constrain the models with Cosmic Chronometers (CC), Dark Energy Spectroscopic Instrument baryon acoustic oscillations (data release 2), CMB priors, and a Gaussian likelihood for the growth index. We show that the constant $\gamma$ approximation is accurate for the parametrization linked to the energy density parameter of dark energy,  but presents a non-negligible error for the other case, which we treat as a systematic error in the analysis. We find a  $2.4\sigma-3.4\sigma$  tension level with constant $G_{\rm eff}$, depending on the parametrization. The results indicate that $G_{\rm eff}<G_N$ around the period of accelerated expansion, corresponding to a weaker effective gravitational interaction on cosmological scales, which leads to a suppression of the growth of cosmological structures.
\end{abstract}
\maketitle

\section{Introduction}
The study of structure formation in the Universe is an important tool for understanding its evolution, energy content, and testing models that can explain the recent accelerated expansion. A particular simple but powerful method is the study of the linear growth rate of matter density perturbations \cite{Linder:2005in,Amendola2013,Huterer2018}, given by
\begin{equation}
f(a) \equiv \frac{d\ln \delta_{m}(a)}{d\ln a},
\end{equation}
where $a$ is the scale factor and $\delta_m(a)=\delta\rho _m (a)/\bar{\rho}_m(a)$  the matter density contrast, which is scale independent at late times for a great variety of cosmological models. It is well known that, in many models, the growth rate can be accurately parametrized by
\begin{equation}
f(a)\simeq\Omega_{m}(a)^{\gamma},\label{eq:growth_rate-gamma}\,
\end{equation}
where $\gamma$  is the growth index, usually assumed constant,  $\Omega_{m}(a)=\bar{\rho}_m(a)/\bar{\rho}_c(a)$  is the matter density parameter and $\bar{\rho}_c(a)$ the critical energy density. For the $\Lambda$CDM model, we have $\gamma\simeq0.55$ \citep{Wang:1998gt,Linder:2005in}. More precisely, constraining the background only with Cosmic Chronometers (CC) data,  The companion paper, \citep{Cortes2025}, found $\gamma=0.551\pm0.001$ for the $\Lambda$CDM model, with parametrization errors about $0.2\%$.

However, a recent determination of $\gamma$ based on the observation of baryon acoustic oscillations (BAO), supernova type Ia (SNIa), cosmic microwave background (CMB), redshift space distortions and galaxy correlations indicated $\gamma=0.633^{+0.025}_{-0.024}$ \citep{Nguyen:2023fip}, which represents a tension of $3.7\sigma$  with respect to the $\Lambda$CDM value, implying a growth suppression with respect to the standard model ($\Lambda$CDM).  Later, it was shown that neither homogeneous nor clustering dark energy (DE) models with equation of state (EoS) described by the $w_0w_a$  parametrization can naturally provide such high $\gamma$ values, and no significant modifications in the determination of $\gamma$ are expected for this EOS parametrization \citep{Cortes2025}.

In Ref.   \citep{Nguyen:2023fip} it is was also discussed that the measured $\gamma$ value solves the $S_{8}$  tension \citep{DiValentino2021}, where   $S_{8}\equiv\sigma_{8}\sqrt{\Omega_{m0}/0.3}$,  $\sigma_{8}$ denotes the rms amplitude of linear matter fluctuations smoothed on $8\,h^{-1}\,\mathrm{Mpc}$ and $\Omega_{m0}$ is the present-day matter density parameter. Nevertheless, it is important to note that the most recent KIDS analysis indicated a much smaller $S_8$ tension \citep{Wright:2025xka}.  This result suggests that an analysis with the latest weak lensing data might reduce the value of $\gamma$.  Moreover, it was argued that the dependence of $\sigma_8$ on $h^{-1}{\rm Mpc}$ units exacerbates the growth tension when dealing with galaxy clustering data \citep{Sanchez:2020vvb,Forconi2025}. These papers suggest that a more robust quantity would be $\sigma_{12}$, the analogous of $\sigma_8$  but computed on a $12{\rm Mpc}$ scale.

Considering the possibility that $\gamma$  has a value significantly higher than the $\Lambda$CDM prediction and that even more general DE models have substantial difficulties in explaining it, one can ask what other cosmological models could naturally provide high $\gamma$  values. In this paper, we study the possibility that Newton's gravitational constant, $G_N$,  can change over cosmological times and decrease the growth of matter perturbations, thus increasing $\gamma$.

Given the growth suppression implied by $\gamma>0.55$ with respect to the $\Lambda$CDM model,  $G_{\rm eff}<G_N$  is expected. In an analysis using redshift space distortion data, Ref. \cite{Toda:2024fgv},  found $\mu_1 <1$ at $2.6-2.8\sigma$ confidence level (CL), where $\mu_1$  can be interpreted as a constant  $G_{\rm eff}/G_N$ at $z<1$.

In this work, we show that the constant $\gamma$ parametrization accurately describes two popular time-dependent $G_{\rm eff}$ parametrizations and that there is a strong correlation between $\gamma$ parameters describing  $G_{\rm eff}$.  We also show that the measured $\gamma$ value leads to constraints that disfavor the standard model with mild significance ($2.4-3.4\sigma$), depending on the parametrization considered,  implying a lower $G_{\rm eff}$ in the recent past and now.

Our analysis is as follows: we assume that the background evolution is described by the $\Lambda$CDM model and make use of two datasets to constrain the background parameter $\Omega_{m0}$:  CC data and Dark Energy Spectroscopic Instrument Data Release 2 (DESI DR2) BAO data plus CMB priors (DESI+CMB). Given that the reported $\gamma$ measurement was done assuming the $\Lambda$CDM model, we also include a Gaussian likelihood for this measurement in both cases.  For the CC data case, we get very loose constraints on $\Omega _{m0}$ that are used to test the accuracy of the constant $\gamma$ parametrization for two parametrizations describing the effective Newton's constant,  $G_{\rm eff}$, on cosmological times. When using DESI+CMB data, we get much tighter constraints on $\Omega _{m0}$; however, the corresponding uncertainties on the parameters describing $G_{\rm eff}$  are very similar to the CC case, indicating a strong correlation between $\gamma$  and  $G_{\rm eff}$.

Many alternative gravity theories can present a time-varying  $G_{\rm eff}$, while still evading Solar System constraints; see Ref. \citep{Uzan:2024ded} for a recent discussion. Such a scenario is viable when the underlying models incorporate screening mechanisms. These mechanisms preserve the locally measured Newton’s constant, suppressing deviations from General Relativity (GR) in high-density environments (e.g. terrestrial or Solar System experiments), while allowing modifications to emerge on astrophysical and cosmological scales. Representative screening mechanisms include: (i) environment-dependent screening — chameleon/$f(R)$, symmetron — where the scalar field effectively decouples in dense regions, recovering GR in the appropriate limit \citep{Khoury2003,Hu2007,Hinterbichler2010}; (ii) derivative (Vainshtein) screening in Dvali-Gabadadze-Porrati (DGP)/Galileon, in which nonlinear derivative terms screen the scalar field in the vicinity of dense sources \citep{Vainshtein1972,Koyama2005,Nicolis2008}; and (iii) kinetic/disformal realizations (e.g., K-mouflage and effective field theory (EFT) in beyond-Horndeski theories), in which the effective coupling in the Poisson equation becomes a scale-, time-, and environment-dependent quantity \citep{Brax2014,Bellini2014}.

In this paper, we remain agnostic about the fundamental model producing time variations on $G_N$ and adopt a purely phenomenological approach, considering two parametrizations for $G_{\rm eff}$. This parametrized quantity encodes the net cosmological deviation from GR that affects the growth of matter perturbations on cosmological scales, under the assumption that local constraints are satisfied via screening.

This paper is structured as follows: in Sect. \ref{sec:Cosmology} we discuss the assumptions for the background cosmology, the growth of matter perturbations, and the parametrizations for $G_{\rm eff}$. Later, in Sect. \ref{sec:Bayesian} we present the data and methodology employed, and a discussion about the accuracy of the growth index parametrization for the parametrization used. Finally, we compile the sampling results in Sect. \ref{sec:Results} and summarize the conclusions in Sect. \ref{sec:Conclusion}.

\section{Cosmological Models\protect\label{sec:Cosmology}}

When dealing with modified gravity models, it is possible to use an effective separation between the background evolution and the impact of GR modifications on the evolution of cosmological perturbations, as discussed in Refs. \cite{Skordis:2008vt,Lombriser:2014ira,Bellini2014}, for instance. Although this approach is theoretically motivated and common practice in cosmological data analysis, specific models might fail to describe background evolution and be consistent with astrophysical constraints, e.g., Refs. \cite{Sakstein:2017xjx,Ezquiaga:2017ekz,Creminelli:2017sry}.

In this context, our analysis is valid for modified gravity models that can provide a background evolution close to the $\Lambda$CDM one, which in turn allows us to use the $\gamma$  measurement to constrain $G_{\rm eff}$.  Then we can define a standard flat $\Lambda$CDM background cosmology following GR with the Friedmann-Lemaître-Robertson-Walker metric,
\begin{equation}
ds^{2}=-c^{2}dt^{2}+a(t)^{2}\left[dr^{2}+r^{2}d\Omega^{2}\right]\, .
\end{equation}
In which case, we have the Friedmann equation for the $\Lambda$CDM model
\begin{equation}
H^{2}=\left(\frac{\dot{a}}{a}\right)^{2}=H_{0}^{2}\left(\Omega_{m0}a^{-3}+\Omega_{\Lambda}\right),
\end{equation}
where  $\Omega_{\Lambda}=1-\Omega_{m0}$ is the DE density parameter associated with the cosmological constant at the present time. With those considerations, the background cosmology at late times is characterized by two parameters: $(h,\Omega_{m0})$, where $h\equiv H_{0}/\left(100\text{km s}^{-1}\text{Mpc}^{-1}\right)$.

\subsection*{Linear growth of matter perturbations}

Before proceeding with the analysis for the impact of $G_{\rm eff}$ on the growth of matter perturbation, it is clarifying to make a brief review of the most usual parametrization for MG. Following Ref. \cite{Koyama:2015vza}, we consider the line element
\begin{equation}
    ds^2= -(1+2\Psi) dt^2 + a^2(t)(1-2\Phi)d\vec{x}^2\,.
\end{equation}
In Fourier space and on subhorizon scales, the linear evolution of matter density contrast is given by
\begin{equation}
    \ddot{\delta}_m+2H\dot{\delta}_m=-\frac{k^2}{a^2}\Psi\,.
    \label{eq:growth-1}
\end{equation}
Modifications of GR in cosmology are usually parametrized by $\mu=\mu(a,k)$ and $\Sigma=\Sigma(a,k)$, as follows:
\begin{equation}
k^2\Psi=-4\pi G \mu\bar{\rho} \Delta
\label{eq:psi-parametrization}
\end{equation}
and
\begin{equation}
    k^2(\Phi +\Psi)=-8\pi G\Sigma \bar{\rho}\Delta\,,
\end{equation}
where $\bar{\rho}\Delta=\sum_i\bar{\rho}_i\Delta_i $  is the total density perturbation in the rest frame; $\Delta_i = \delta_i+ 3aHk^{-2}(1+w_i)\theta_i $, $\delta_i$  is the density contrast, $w_i$  the equation of state parameter and $\theta_i$ the divergent of  the fluid peculiar velocity for each fluid species $i$. Therefore, the growth of matter perturbations directly depends on the function $\mu$. Consequently, non-standard values of $\gamma$ will be associated with $\mu\neq1$. Departures from the usual lensing potential, parametrized by $\Sigma=\mu(\eta+1)/2$, also depend on $\mu$ and additionally on the difference between the gravitational potentials parametrized by $\eta = \Phi/\Psi$.  In GR, we have $\mu=\eta=1$, and for the $\Lambda$CDM background, we must have $\gamma=0.55$.  In general, MG theories will have complex forms for $\mu$ and $\Sigma$ as a function of the scalar field responsible for the modifications with respect to GR.

Equation  (\ref{eq:growth-1}) can be rewritten using Eq. (\ref{eq:psi-parametrization})  on small scales and identifying $\mu=\mu(a)$  as $G_{\rm eff}(a)/G_N$, yielding   \citep{Nesseris2007,Tsujikawa2009,Nesseris2011}
\begin{equation}
\delta^{\prime\prime}_m+\frac{3}{2}\frac{\Omega_{m}(a)}{a}\delta^{\prime}_m-\frac{3}{2}\frac{G_{\rm eff}(a)}{G_N}\frac{\Omega_{m}(a)}{a^{2}}\delta_m=0\,,
\label{eq:delta-G-EDO}
\end{equation}
where the prime denotes derivatives with respect to the scale factor,  and $\Omega_{m}(a)$ is the time-dependent density parameter associated with the baryonic and cold dark matter components. Note that $\Omega_{m}$ excludes the contribution of massive neutrinos. As we will show, large variations on  $\Omega_{m0}$ produce much smaller changes on $\gamma$. Consequently, a shift of $10^{-4}$ order on  $ \Omega_{m0}$, which is the usual impact associated with massive neutrinos, has a negligible impact on $\gamma$.

Solving Eq. \eqref{eq:delta-G-EDO} numerically allows us to compute the linear growth rate $f= d\ln\delta_m/d\ln a$ and fit a constant $\gamma$ in the parametrization $
f=\Omega_{m}^{\gamma}(a)$. To set the initial conditions, we assume that at $z=100$ the Universe is matter-dominated and $G_{\rm eff}=G_N$. Thus the usual Einstein-de Sitter solution holds initially, $\delta_m(a)\propto a$. Our code is publicly available at \url{https://github.com/icarob-eng/gimbal_pub}.

For a time-varying $G_{\rm eff}$, the growth of matter will be enhanced when  $G_{\rm eff}>G_N$, yielding $\gamma<0.55$. When $G_{\rm eff}<G_N$, the growth is suppressed, and we must have  $\gamma>0.55$. Next, we present the two parametrizations for $G_{\rm eff}$ considered in our work.

\subsection*{Parametrizations for $G_{\rm eff}$}

We consider two parametrizations for effective gravitational coupling,  $G_{\rm eff}(a)$. The first one is based on a second-order Taylor expansion of $G_{\rm eff}(a)$ \citep{Nesseris2011}, as follows:
\begin{equation}
\frac{G_{\rm eff}(a)}{G_N} = g_{0}+g_{1}\left(a-1\right)+\frac{g_{2}}{2}\left(a-1\right)^{2}.
\end{equation}
In this form,  $g_{0},\,g_{1}$ and $g_{2}$ are dimensionless model parameters. As explained in Refs. \cite{Nesseris2007,Nesseris2011},  when considering the Solar System constraints, which impose strong constraints on the local time variation of $G_N, $ one has to assume that $G_{\rm eff}(a)$ is a slowly varying function, thus  $g_1=0$. Moreover,  according to Big Bang Nucleosynthesis (BBN) constraints \cite{Bambi2005,Uzan:2024ded}, we must have  $G_{\rm eff}\simeq G_N$ at BBN time.  Under these assumptions, we have
\begin{equation}
\frac{G_{\rm eff}(a)}{G_N}=1+\frac{g_{2}}{2}\left(\left(a-1\right)^{2}-1\right).\label{eq:Geff-g2}
\end{equation}
Note that the assumption of slowly varying $G_{\rm eff}$ is more restrictive than assuming screening, in which case  $G_{\rm eff}$ could have a more substantial time variation on cosmological scales without affecting local measurements of  $G_{\rm eff}$.  In this model, $g_{2}>0$ is associated with a decaying $G_{\rm eff}$ at low $z$, which is the expected behavior according to the current growth index measurements.

The second parametrization associates the variation of $G_{\rm eff}$  directly with DE energy density, \citep{Ferreira:2010sz}, and was used in the recent DESI Full Shape Analysis \citep{DESICollaboration2024}:
\begin{equation}
\frac{G_{\rm eff}(a)}{G_N}=\mu(a)\equiv1+\mu_{0}\frac{\Omega_{\Lambda}(a)}{\Omega_{\Lambda}},
\label{eq:Geff-gde}
\end{equation}
where $\mu_{0}$ is a dimensionless free parameter and $\Omega_{\Lambda}(a)$ is the time-dependent DE density relative to the critical density.  In this case, $G_{\rm eff}$  also tends to $G_N$ at high $z$ and $\mu_0<0$ is associated with $\gamma >0.55$. As the results will show, this parametrization allows for a stronger time variation.

\section{Bayesian Analysis\protect\label{sec:Bayesian}}

To constrain the background cosmology using the Monte Carlo Markov Chain (MCMC) method, we use the CC or DESI+CMB datasets with a Gaussian likelihood based on the growth index measurement. When using CC data, we are interested in determining a large but meaningful space parameter to test the accuracy of the constant $\gamma$ parametrization. When using the DESI+CMB data set, we look for a better determination of $\Omega_{m0}$ and its impact on $G_{\rm eff}$.  As we will show, this better determination of the background has small impact on the determination of  $G_{\rm eff}$. Therefore, our results should have no significant modifications when using other background data combinations.

The CC likelihood uses a set of 32 cosmic chronometers $H(z)$ data
compiled by Ref. \citep{Favale2023}, with data from Refs. \citep{Moresco2012,Moresco2015,Moresco2016,Zhang2014,Jimenez2003,Simon2005,Ratsimbazafy2017,Stern2010,Borghi2022}. The particular composition of the likelihood function
\begin{equation}
L_{\text{CC}}(D|\theta)\propto\exp\left(-\frac{\chi_{CC}}{2}^{2}\right)
\end{equation}
is the same as described in the data section of Ref. \citep{Cortes2025}, with 17 uncorrelated data points provided by Refs. \citep{Zhang2014,Jimenez2003,Simon2005,Ratsimbazafy2017,Stern2010,Borghi2022}
and 15 data points from Refs. \citep{Moresco2012,Moresco2015,Moresco2016}
to which the systematic errors were considered with a covariance
matrix for the BC3 model as described at \url{https://gitlab.com/mmoresco/CCcovariance}.

The DESI likelihood we use, $L_{\text{DESI}}(D|\theta)$,  is the one implemented in \texttt{COBAYA} \citep{Torrado2020} for DESI BAO DR2 \citep{DESICollaboration2025}. In this case, we also use correlated priors on $\left(\theta_{*},\omega_{b},\omega_{bc}\right)_{\text{CMB}}$ obtained from CMB analysis, as
described in Appendix A of Ref. \citep{DESICollaboration2025}. Here
$\theta_{*}$ is the angular scale of CMB acoustic peaks, $\omega_{b}\equiv\Omega_{b0}h^{2}$
is the present-time baryonic physical density parameter,  $\omega_{bc}=(\Omega_{b0}+\Omega_{c0})h^{2}$
is the present-time baryonic plus cold dark matter physical density parameter. We also assume massive neutrinos with $m=0.06eV$. We denote this set of priors as $P(\theta)_{CMB}$.  We make use of \texttt{CAMB} \citep{Lewis2000,Howlett2012} to calculate the BAO observables. For this analysis, we assume that $\Omega_{m0}=(\omega_{bc0})/h^2$ .

The $\gamma$  likelihood consists of a Gaussian distribution around the observed value, in the form
\begin{equation}
L_{\gamma}(D_{\gamma}|\theta)\propto\exp\left(-\frac{1}{2}\left(\frac{\gamma\left(\theta\right)-\gamma_{\rm obs}}{\sigma_{\gamma}}\right)^{2}\right),
\label{gamma-liki}
\end{equation}
where $\gamma(\theta)$ is the growth index for a particular model represented by the set of parameters $\theta=\left(h,\Omega_{m0},g_{2}\right)$
or $\left(h,\Omega_{m0},\mu_{0}\right)$, $\gamma_{\rm obs}=0.633$ and  $\sigma_{\gamma}=0.025$,  as obtained in Ref. \citep{Nguyen:2023fip}. It is important to note that this determination of $\gamma$ depends on the growth of matter perturbations and lensing of photons, via CMB lensing and weak lensing observations included in the analysis. Our study simulates how this measurement constrains  $G_{\rm eff}$  via its impact on the growth of matter perturbations alone. In analyses such as Ref. \citep{DESICollaboration2024}, two free parameters describe GR modifications, $\mu$ and $\Sigma$. Although $\mu\propto G_{\rm eff}$, the constraints in Ref. \citep{DESICollaboration2024}, which indicate that both $\mu$ and $\Sigma$ are compatible with GR, can not be directly compared with our results because it considers one extra parameter.

The combination of CMB priors and $\gamma$ measurement deserves special consideration because, at first sight,  it might indicate a double counting of observables because the analysis in Ref. \cite{Nguyen:2023fip}  uses CMB data to determine $\gamma$. However, as explained in Ref. \cite{Lemos:2023xhs}, the CMB priors are derived in order to eliminate the dependence of CMB on the late time physics, and the dependence of CMB observables on  $\gamma$  is implemented only via corrections on the Halo Model that will affect the lensing potential, which is a late-time effect. Therefore, by construction, the CMB priors should have negligible dependence on $\gamma$.

The subsequent posteriors combine the growth index measurement with the likelihood of the CC data in the form
\begin{equation}
\mathcal{L}(\theta|D)\propto P(\theta)\cdot L_{\text{CC}}(D|\theta)\cdot L_{\gamma}(D_{\gamma}|\theta)
\end{equation}
or the DESI+CMB data, as
\[
\mathcal{L}(\theta|D)\propto P(\theta)\cdot P_{\rm CMB}(\theta) \cdot L_{\text{DESI}}(D|\theta)\cdot L_{\gamma}(D_{\gamma}|\theta)
\]
where $P(\theta)$ represents the baseline prior, as listed in Tab. \ref{tab:List-of-priors}, and $P_{\rm CMB}(\theta)$ represents the CMB correlated priors.

\begin{table}
\caption{\protect\label{tab:List-of-priors}List priors for the basic parameters. For the analysis with the DESI+CMB dataset, the extra parameters $\left(\theta_{*},\omega_{b},\omega_{bc}\right)_{\text{CMB}}$ are
subjected to correlated a Gaussian prior, as expressed in Appendix A of Ref. \citep{DESICollaboration2025} .}

\centering{}%
\begin{tabular}{cc}
\hline
Parameters & Priors \tabularnewline
\hline
 $h$           & $\mathcal{U}(0.5,1)$ \tabularnewline
 $\Omega_{m0}$ & $\mathcal{U}(0.01,0.99)$ \tabularnewline
$g_{2}$ or $\mu_{0}$ & $\mathcal{U}(-1,1)$\tabularnewline
\hline
\end{tabular}
\end{table}

The MCMC sampling was performed using the \texttt{COBAYA} \citep{Torrado2020,Lewis2002}
package with the Gelman-Rubin statistic satisfying $R-1<0.01$ as
a convergence criterion, yielding 2100 to 3100 steps for CC samplings
and 4600 to 5700 steps for DESI+CMB samplings with an acceptance fraction
of $\sim40\%$.  As a post-processing step, we discarded the first $30\%$ samples as a ``burn-in''.

\subsection*{Accuracy of the growth index for the $G_{\rm eff}$ models}

\begin{figure}[h]
\begin{centering}
\includegraphics[scale=0.55]{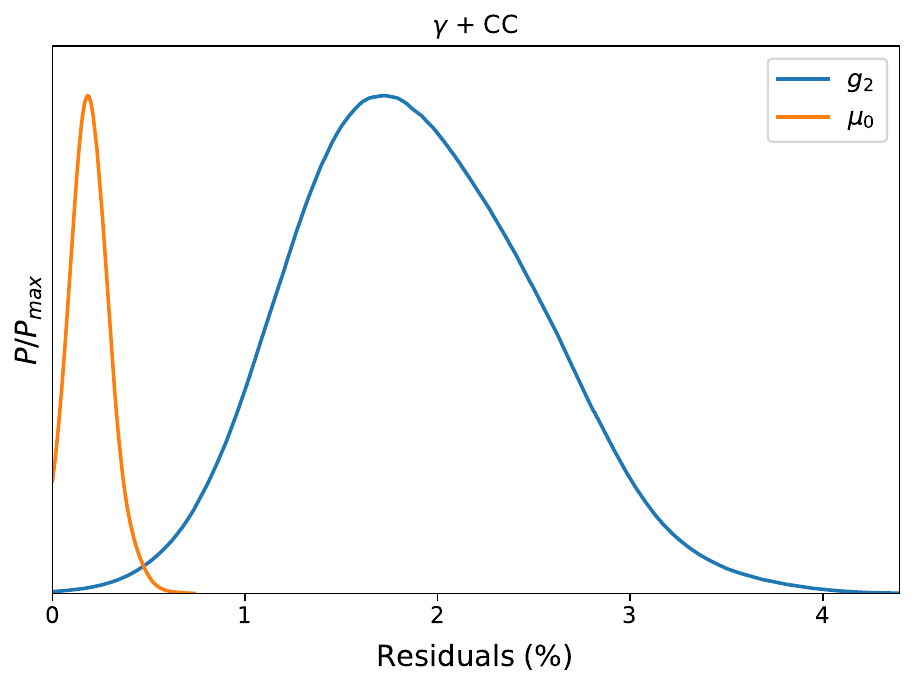}
\par\end{centering}
\caption{\protect\label{fig:Resids}Distribution of rms residuals for the constant growth index parametrization, with respect to $f$, considering different models of modified gravity under CC data. The distributions are normalized by the maximum value.}
\end{figure}

Before turning our attention to the constraints on $g_2$ and $\mu_0$, we must evaluate whether cosmological models based on these parametrizations for $G_{\rm eff}$  can be accurately described by a constant $\gamma$. To analyze this, we run an MCMC sampling with the CC+$\gamma$ likelihoods and compute the root mean square (rms) percent residuals of Eq. \eqref{eq:growth_rate-gamma}, defined for a particular realization as
\begin{equation}
r=100\times \sqrt{\frac{1}{10}\sum_z\left(1-\frac{\Omega_m^{\gamma(\theta)}(z)}{f_\text{num}(z;\theta)}\right)^2}\,[\%],
\label{eq:resid}
\end{equation}
where $f_\text{num}(z;\theta)$ is the linear growth rate for the numerical solution of Eq. \eqref{eq:delta-G-EDO} for the realization with parameters $\theta$,  $\gamma(\theta)$ is the growth index for this realization, and the redshifts $z$ are sampled in 10 evenly spaced points in the interval $0\le z\le2$. The distribution of the rms residuals for the realizations is shown in Figure \ref{fig:Resids}, which was produced with \texttt{GetDist} \citep{Lewis2019}. As the figure shows, the residuals for the $g_{2}$ model are considerable ($r\sim1.89_{-0.70}^{+0.57}[\%]$), while for the $\mu_{0}$ models, most of the residuals are less than half a percent ($r\sim0.200_{-0.011}^{+0.086}[\%]$).

In order to account for the non-negligible residuals in the growth index for the $g_2$ model, we consider it as a systematic error of $3\%\times\gamma_{\rm obs}\simeq 0.019$ summed in quadrature with the $\gamma$ measured uncertainty, which yields
\begin{equation}
\sigma_{\gamma -g_2}=0.031\,.
\label{eq:sigma-resid}
\end{equation}

In the analysis for $g_2$, we consider $\sigma_{\gamma -g_2}$ as the effective uncertainty for the $\gamma$ likelihood, Eq. \eqref{gamma-liki}. For the $\mu_0$ model, the systematic error due to the constant $\gamma$ parametrization is much smaller than the measured $\gamma$  uncertainty and can be safely neglected.

\section{Results\protect\label{sec:Results}}

\begin{figure*}
\begin{centering}
\includegraphics[scale=0.7]{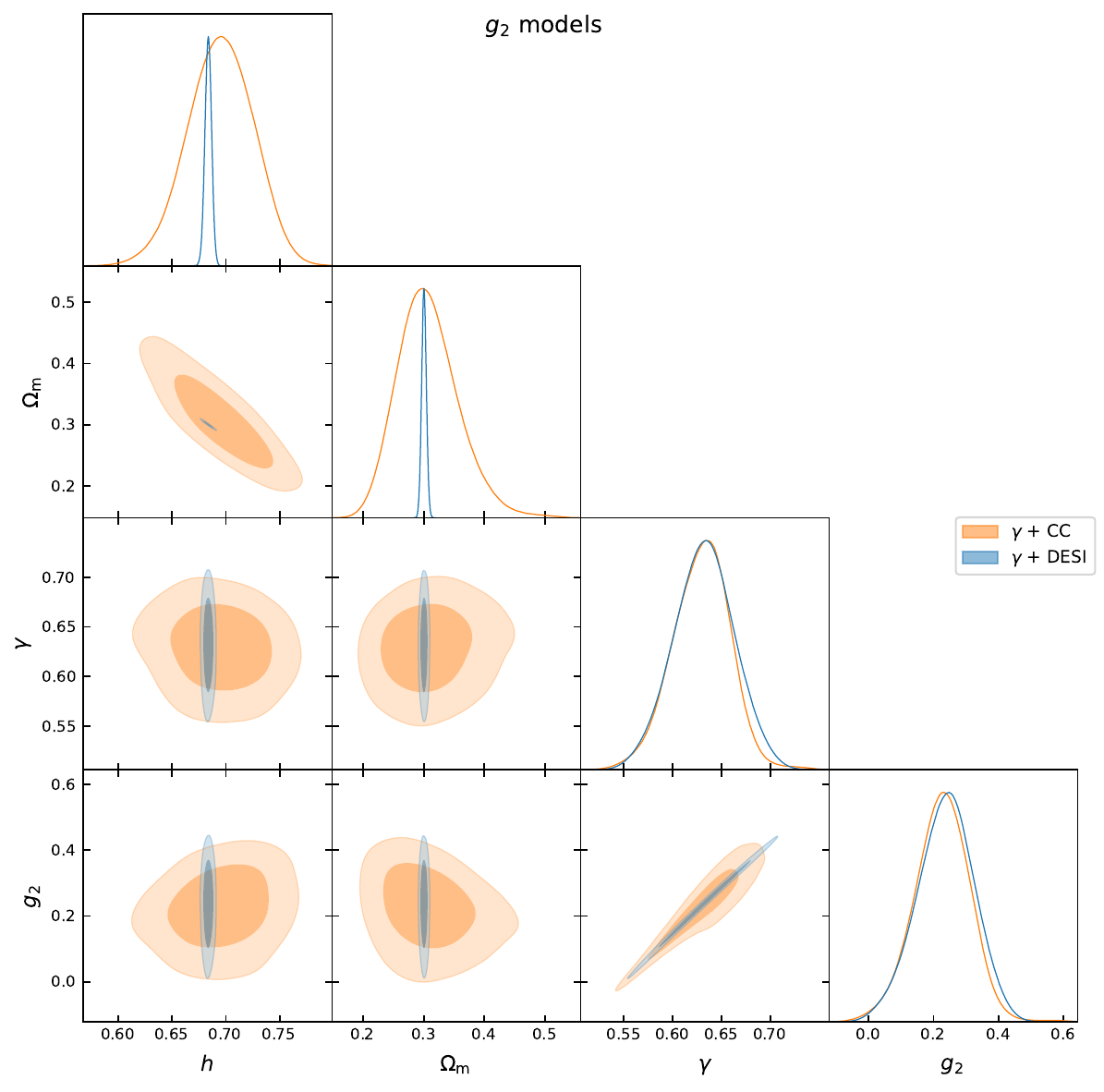}
\par\end{centering}
\caption{\protect\label{fig:trig-g2}Posteriors for the $g_{2}$ model with
the measured value of the growth index considering residuals in the
parametrization \eqref{eq:growth_rate-gamma} ($\gamma=0.633\pm0.031$)
together with either the CC dataset, in orange, or the DESI+CMB dataset, in blue.}

\end{figure*}

\begin{figure*}
\begin{centering}
\includegraphics[scale=0.7]{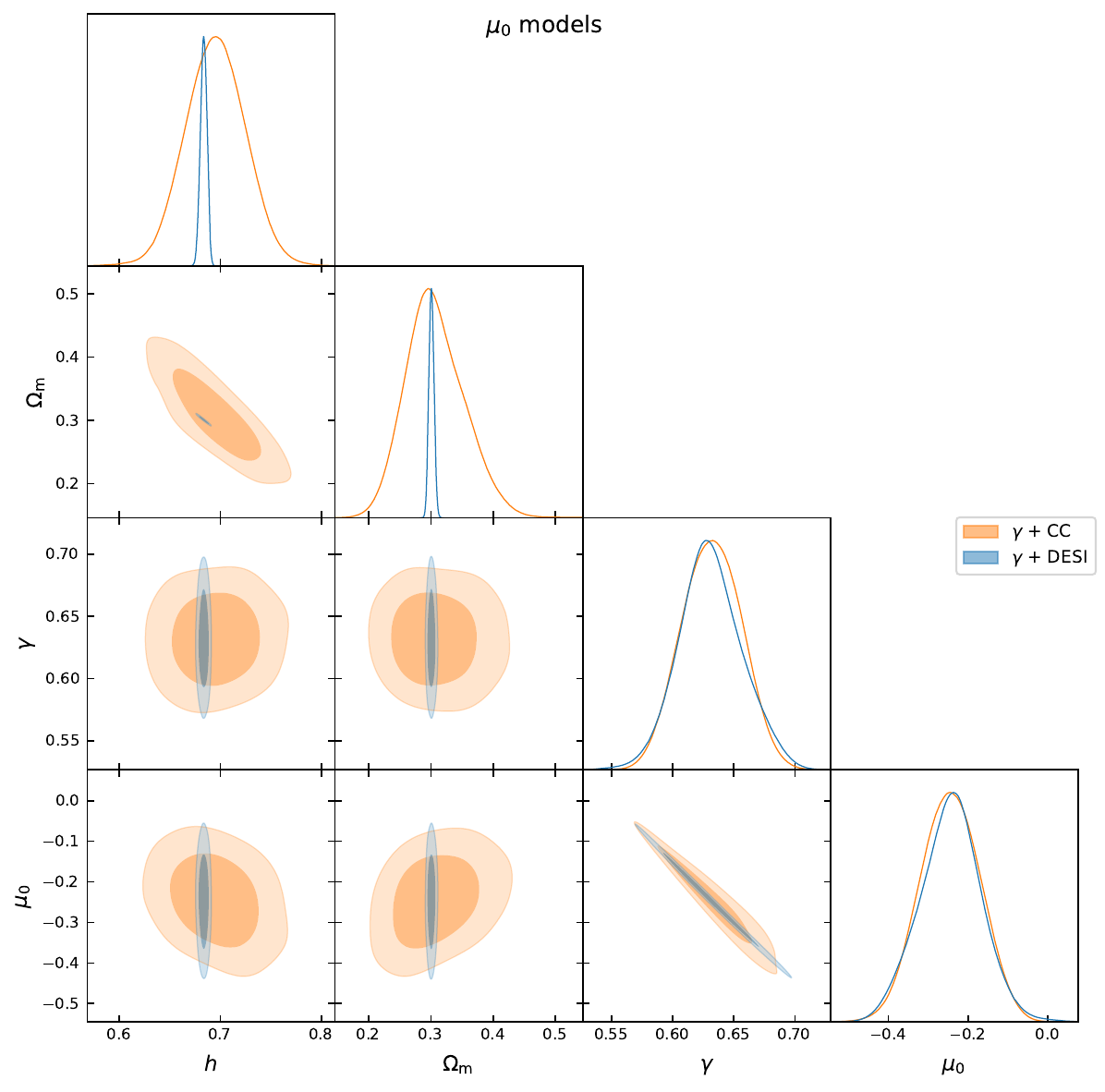}
\par\end{centering}
\caption{\protect\label{fig:trig-mu0}Posteriors for the $\mu_{0}$ model with
the measured value of the growth index ($\gamma=0.633\pm0.025$) together
with either the CC dataset, in orange, or the
DESI+CMB dataset, in blue.}
\end{figure*}

Based on the sampling method described in Sec. \ref{sec:Bayesian},
we obtained the posterior distributions for the $g_{2}$ model as presented in
Figure \ref{fig:trig-g2} and for the $\mu_{0}$ model in Fig. \ref{fig:trig-mu0}. As expected, the DESI+CMB data provide a much more precise determination of background parameters than CC data. However, the corresponding effect in the determination of   $\gamma$ and $g_{2}$ or $\mu_{0}$  is much smaller, demonstrating that variations on the background parameters have a small impact on the determination of $\gamma$. This result also justifies using the $\gamma$ likelihood as independent from the ones for the background parameters.

The main result for both models is a significant preference for time-varying $G_{\rm eff}$. As shown in Figures \ref{fig:trig-g2} and \ref{fig:trig-mu0}, the values $g_2=0$ and $\mu_0=0$ are marginally allowed by the posterior distributions.  There is also a strong correlation between $g_{2}$ or $\mu_{0}$ and $\gamma$. Regarding this correlation, it is clear that the DESI+CMB data diminish the scatter between these parameters with respect to the results based on CC data.

The mean values and marginalized 1D limits for 68\% CL are shown in Table \ref{tab:constraints}, as well as the significance of the tension of the models in relation to the standard constant gravitational model, in terms of  $T_{\sigma}\equiv\overline{g}_{2}/\sigma_{g_{2}}$ or $T_{\sigma}\equiv\overline{\mu}_{0}/\sigma_{\mu_{0}}$. For the computation of the tension in the $g_2$ model, we conservatively assume the largest value of the reported $1\sigma$  limits. Note that the constraints on the model's parameters depend primarily on the accuracy of the growth index's measurement. Consequently, the $g_{2}$ model, which has significant residuals treated as systematic errors, excludes the standard $G_N$  with less significance, namely $2.58\sigma$. On the other hand, the $\mu_{0}$ model, which has negligible residuals in the $\gamma$ fit, allows the determination of $\mu_{0}\neq0$ with greater significance. For the case of DESI+CMB dataset, the tension with $\mu_{0}=0$  is $3.43\sigma$.

\begin{table*}
\caption{\protect\label{tab:constraints}Marginalized 1D constraints for 68\% CL intervals for the parameters of each sampling case and the significance in $T_{\sigma}$ of the exclusion of the constant gravitational model.}

\centering{}%
\begin{tabular}{cccccc}
\hline
Model & Dataset & $h$ & $\Omega_{m0}$ & $g_{2}$ or $\mu_{0}$ & $T_{\sigma}$\tabularnewline
\hline
\hline
\multirow{2}{*}{$g_{2}$} & $\gamma$ + CC & $0.695\pm0.030$ & $0.307_{-0.055}^{+0.042}$ & $0.227_{-0.079}^{+0.088}$ & $2.58$\tabularnewline
 & $\gamma$ + DESI & $0.6837\pm0.0030$ & $0.3003\pm0.0039$ & $0.237_{-0.083}^{+0.092}$ & $2.42$\tabularnewline
\hline
\multirow{2}{*}{$\mu_{0}$} & $\gamma$ + CC & $0.695\pm0.029$ & $0.307_{-0.052}^{+0.041}$ & $-0.245\pm0.074$ & $3.31$\tabularnewline
 & $\gamma$ + DESI & $0.6836\pm0.0030$ & $0.3006\pm0.0039$ & $-0.254\pm0.074$ & $3.43$\tabularnewline
\hline
\end{tabular}
\end{table*}

As a forecast exercise, we consider a hypothetical measurement of  $\gamma=0.633$ with  $\sigma_{\gamma}=0.02$, with the background constrained by DESI+CMB data. The Euclid mission expects to measure the growth index with accuracy below $0.02$, \citep{EuclidCollaboration2024}. In this case, the tension in reproducing $G_N$ is higher than $3.5\sigma$ for the $g_2$ model and $4.1\sigma$  for the $\mu_0$ model. Therefore, near-future measurements of $\gamma$ might be decisive in indicating a failure of GR on cosmological scales and late times.

\subsection{Fitting the correlation and $G_{\rm eff}$ behavior}

A can be seen in Figs. \ref{fig:trig-g2} and \ref{fig:trig-mu0},  there is a linear relation between the growth index  and $g_{2}$ or  $\mu_{0}$.  A linear regression based on the DESI+CMB posteriors produced the relations
\begin{equation}
\gamma\left(g_{2}\right)=0.35g_{2}+\gamma_{\Lambda}
\label{eq:reg-g2}
\end{equation}
 and
\begin{equation}
\gamma(\mu_{0})=-0.34\mu_{0}+\gamma_{\Lambda},
\label{eq:reg-mu0}
\end{equation}
where $\gamma_{\Lambda}\equiv0.55$. The residuals of these fits concerning the sampled distributions are less than $0.6\%$ in both cases.

We checked that the fits in Eqs. \eqref{eq:reg-g2} and \eqref{eq:reg-mu0} have accuracy better than $3\%$ if we set the $\gamma$  value in its likelihood at the $\Lambda$CDM one with $\sigma _{\gamma}=0.04$, still constraining the background parameters with DESI+CMB data. Therefore, these fits can used for   $0.29<\Omega_{m0}<0.31$ and $0.47<\gamma<0.67$, roughly the $3\sigma$ interval.

\begin{figure}
    \centering
    \includegraphics[width=1\linewidth]{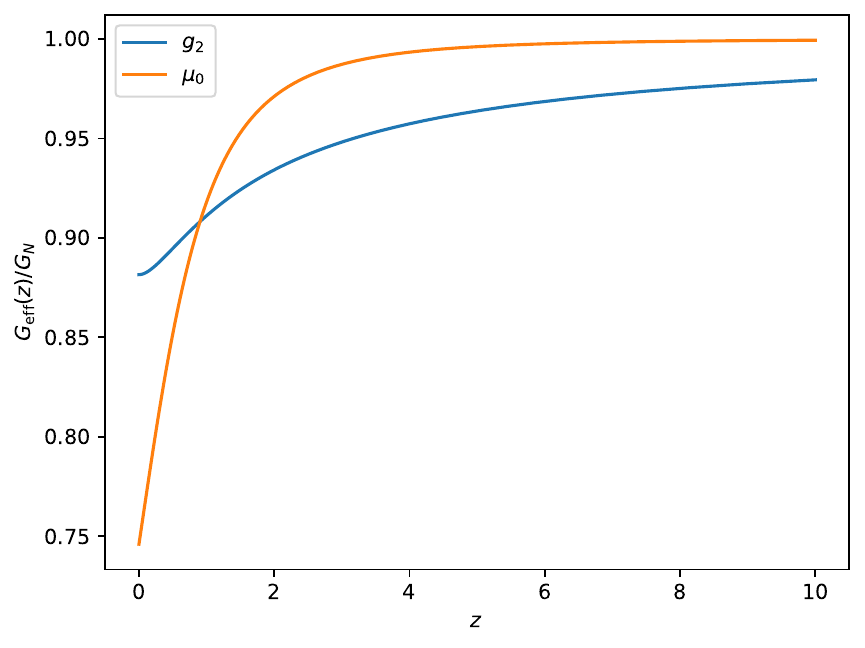}
    \caption{Variation of  $G_{\rm eff}$ for the two parametrizations considered as a function of redshift.}
    \label{fig:placeholder}
\end{figure}

Considering the mean values of $g_2$ and $\mu_0$ in Tab. \ref{tab:constraints} for the DESI+CMB analysis, we show the evolution of $G_{\rm eff}$ in Figure \ref{fig:placeholder}. As expected from its construction, in the $g_2$ parametrization, $G_{\rm eff}$ has a mild time variation, whereas for the  $\mu_0$ case  $G_{\rm eff}$ has a much stronger decay at low redshift. Since  $\mu_0$ has a faster decay,  $G_{\rm eff}$ approaches $G_N$ much earlier than $g_2$.

\section{Conclusions\protect\label{sec:Conclusion}}

Throughout this paper, we investigated the consequences of the growth index measurement of Ref. \citep{Nguyen:2023fip} for two parametrizations of the time variation of Newton’s gravitational constant, described in Eqs. \eqref{eq:Geff-g2} and \eqref{eq:Geff-gde}, namely the $g_{2}$ and $\mu_{0}$ models.

We first analyzed whether the constant $\gamma$ parametrization for the growth rate can accurately describe the models. For this study, we constrained the background evolution only with CC data, which provide loose constraints on the background parameters, thus allowing a more general analysis to evaluate the error associated with the constant $\gamma$ parametrization.  We found that a relevant fraction of the realizations of the $g_{2}$ model present significant $3\%$ residuals in the growth index parametrization of Eq. \eqref{eq:growth_rate-gamma} (see Fig. \ref{fig:Resids}), which motivated the inclusion of a systematic uncertainty in the $\gamma$ likelihood, Eq. \eqref{eq:sigma-resid}. For the $\mu_0$ case, the typical residuals are $0.2\%$ and can be neglected.

The main analysis also considered DESI BAO DR2 data plus CMB priors on background parameters, including a Gaussian likelihood centered on the reported growth index. The resulting posterior distributions for both analyses (Figs. \ref{fig:trig-g2} and \ref{fig:trig-mu0}) show that background constraints have little impact on the bounds of $\gamma$, $g_2$, or $\mu_0$. Moreover, there is a linear relation between the growth index and the modified-gravity parameters, which for the DESI+CMB posteriors can be fitted with low residuals, as illustrated in Eqs. \eqref{eq:reg-g2} and \eqref{eq:reg-mu0}.

Most importantly, both models reveal a tension exceeding $2.4\sigma$ concerning the standard gravitational constant, as summarized in Table \ref{tab:constraints}. The $\mu_{0}$ case, in particular, reaches up to $3.4\sigma$ of tension. Notably, the constraints tend toward $g_{2}>0$ and $\mu_{0}<0$, which both imply $G_{\rm eff}/G_N<1$, with values varying around $G_{\rm eff}(a=1)/G_N\simeq0.76$ for $\mu_0$ and $G_{\rm eff}(a=1)/G_N\simeq0.88$ for $g_2$. This corresponds to a weaker effective gravitational interaction on cosmological scales, suppressing the structure-formation process with respect to GR. Such a behavior is consistent with derivative screening of the Vainshtein type, as realized in quartic Galileon and in the self-accelerating branch of the DGP model \citep{Li2013,Li2013a}. However, the quartic Galileon was already shown to be incompatible with the late background expansion and astrophysical constraints \cite{Sakstein:2017xjx}.

If future analysis of upcoming data still indicates a high $\gamma$ value, but with smaller uncertainty, it can be considered as evidence of Modified Gravity. The Euclid mission expects to measure $\gamma$ with uncertainty below $0.02$. If the central value of this expected measurement is close to or higher than the current one, $0.633$, the tension with GR can be higher than $3.5\sigma$ and $4.1\sigma$ for the $g_2$ and $\mu_0$  models, respectively.

\section*{Acknowledgments}

We thank Miguel Quartin for useful discussions. LGM thanks \textit{Conselho Nacional de Desenvolvimento Científico e Tecnológico} (CNPq, Brazil) for partial financial support—Grant: 307901/2022-0. IBSC thanks the \textit{Conselho Nacional de Desenvolvimento Científico e Tecnológico} for the undergraduate research scholarship N° 02/2025 project PVJ22437-2024 (PIBIC-CNPq) and \textit{Universidade Federal do Rio Grande do Norte} for the undergraduate research scholarship N° 09/2024 (PIBIC-PROPESQ) project PIJ23184-2024.

\bibliographystyle{apsrev4-1}
\bibliography{referencias.bib}

\end{document}